# Quantum-dot single-photon source on a CMOS silicon photonic chip integrated using transfer printing

Ryota Katsumi[1, 3, a)], Yasutomo Ota[2, b)], Alto Osada[2], Takuto Yamaguchi[1], Takeyoshi Tajiri[1], Masahiro Kakuda[2], Satoshi Iwamoto[1,2], Hidefumi Akiyama[3], and Yasuhiko Arakawa[2]

[1] *Institute of Industrial Science, The University of Tokyo, 4-6-1 Komaba, Meguro-ku, Tokyo, Japan*

[2] *Institute for Nano Quantum Information Electronics, The University of Tokyo, 4-6-1 Komaba, Meguro-ku, Tokyo, Japan*

[3] *Institute for Solid State Physics, The University of Tokyo, 5-1-5 Kashiwanoha, Kashiwa, Chiba, Japan*

**Abstract**

**Silicon photonics is a powerful platform for implementing large-scale photonic integrated circuits (PICs), because of its compatibility with mature complementary-metal-oxide-semiconductor (CMOS) technology. Exploiting silicon-based PICs for quantum photonic information processing (or the so-called silicon quantum photonics) provides a promising pathway for large-scale quantum applications. For the development of scalable silicon quantum PICs, a major challenge is integrating on-silicon quantum light sources that deterministically emit single photons. In this regard, the use of epitaxial InAs/GaAs quantum dots (QDs) is a very promising approach, because of their capability of deterministic single-photon emission with high purity and indistinguishability. However, the required hybrid integration is inherently difficult and often lacks the compatibility with CMOS processes. Here, we demonstrate a QD single-photon source (SPS) integrated on a glass-clad silicon photonic waveguide processed by a CMOS foundry. Hybrid integration is performed using transfer printing, which enables us to integrate heterogeneous optical components in a simple pick-and-place manner and thus assemble them after the entire CMOS process is completed. We observe single-photon emission from the integrated QD and its efficient coupling into the silicon waveguide. Our transfer-printing-based approach is fully compatible with CMOS back-end processes, and thus will open the possibility for realizing large-scale quantum PICs that leverage CMOS technology.**

___________________________

a) E-mail: katsumi@iis.u-tokyo.ac.jp.

b) E-mail: ota@iis.u-tokyo.ac.jp.

Photonic integrated circuits (PICs) are powerful platforms for realizing large-scale quantum photonic information processing. Of the various PIC technologies[1], silicon photonics[2–5] is one of the most attractive platforms for implementing large-scale PICs, because of its compatibility with mature complementary-metal-oxide-semiconductor (CMOS) technology[6]. In this paradigm, well-developed silicon nanophotonic elements, such as high-Q ring-resonators, low-loss waveguides, and modulators are available without major development, and thus enable for direct access to system-level research of PICs. Silicon quantum photonics exploits the power of silicon photonics and provides a fascinating route for large-scale photonic quantum information processing. Silicon quantum PICs have already shown their promise for quantum key distribution[7,8], linear optical quantum computation[9,10], and boson sampling[11,12]. However, current silicon quantum photonics inherently lacks scalability due to the probabilistic nature of single-photon sources (SPSs) that have been implemented on silicon, including those based on spontaneous parametric down conversion and spontaneous four wave mixing[13]. Several approaches have been proposed to overcome the limitations of purely silicon-based SPSs[14], but these remain technologically difficult to implement faithfully on silicon.

An alternative approach to implementing deterministic SPSs is the hybrid integration of solid-state quantum emitters, such as those in diamonds, two-dimensional materials, and carbon nanotubes[15]. Among them, InAs/GaAs self-assembled quantum dots (QDs) are highly promising because of their proven potential to deterministically emit single photons with high purity and indistinguishability[16–18]. The QDs can be engineered to emit single photons in the telecom band, where silicon is optically transparent[19–22]. However, their hybrid integration is inherently difficult, as the random position and emission wavelength of each epitaxial QD hinder the deterministic integration of a desired QD on a proper position of the target PIC. The difficulty becomes more pronounced when utilizing conventional heterogeneous integration techniques, such as wafer bonding and direct epitaxial growth[23]. Thus far, there are a few reports on the integration of QD-based SPSs on silicon-based photonic platforms (including those based on $Si_3N_4$)[24–26].

However, none of them has been implemented in such a manner that the entire fabrication process is fully compatible with the current CMOS technology. Pursuing a means of fusing fully CMOS-processed silicon photonics chips with QD SPSs is imperative for leveraging the power of silicon photonics.

In this work, we demonstrate the integration of an InAs/GaAs QD SPS into a silicon photonic waveguide prepared by a CMOS foundry. We utilize transfer printing[27–30] to integrate the SPS in a simple pick-and-place manner onto the silicon waveguide after the entire CMOS process is completed. In design, we show that the transfer-printed SPS supports a near-unity coupling efficiency of QD radiation into the waveguide, which is enabled by the use of a photonic crystal (PhC) nanobeam cavity that facilitates the necessary optical coupling. Experimentally, we demonstrate single-photon generation from the QD integrated on chip and its coupling into the silicon waveguide. Our hybrid integration approach can selectively integrate appropriate SPSs on desired positions of a PIC, thus potentially enabling scalable implementation of multiple identical QD SPSs into a highly functional silicon quantum PIC.

Figure 1(a) shows a schematic of the device structure we investigate in this study. A QD SPS is placed above a CMOS-processed silicon waveguide cladded with glass. This arrangement is well suited for assembly with transfer printing. The QD is embedded in a one-dimensional (1D) PhC nanobeam cavity, which mediates efficient funneling of QD emission into the underneath waveguide. We consider a GaAs-based (refractive index, $n$ = 3.4) nanobeam cavity on a glass clad with a width of 450 nm and thickness of 180 nm. In the nanobeam, air holes with a radius of 78 nm are patterned with a period of 300 nm. The pattern period around the cavity center is modulated, resulting in the formation of a PhC nanobeam cavity[31–33]. When the cavity is not coupled to the underneath waveguide, the fundamental mode of the cavity (resonating at $\lambda$ = 1,170 nm) supports a high Q-factor of $Q = 5.4\times10^6$ and a small mode volume of $V = 0.43(\lambda/n)^3$, which is calculated using a 3D finite difference time domain (FDTD) algorithm. Figure 1(b) shows a cross-sectional schematic of the investigated device. We set the silicon waveguide width and thickness to be 250 nm and 210 nm, respectively. These parameters were chosen to achieve phase-matching between the cavity mode and a transverse electric-like waveguide mode[34]. The optical coupling of the QD into the silicon waveguide can be optimized by tuning the vertical distance between the cavity and waveguide ($d$)[35], which affects the efficiencies of both cavity-waveguide coupling ($\eta$) and emitter-cavity coupling ($\beta$). Figure 1(c) plots dependences of coupling efficiencies and Q-factors with respect to $d$, which is computed using the 3D FDTD method. We observed an increase of $\eta$ when $d$ was reduced from 700 to 400 nm. This increase accompanies an exponential reduction of $Q$, which is induced by the evanescent coupling of cavity mode leakage into the silicon waveguide. The deterioration of $Q$ in turn diminishes $\beta$ (mainly for $d < 450$ nm), which, however, is marginal thanks to a large Purcell effect ($\propto Q/V$) facilitated in the small-$V$ nanocavity. Overall, total coupling efficiency ($\eta\beta$) reaches a maximum value of 99.5% at $d = 450$ nm. It is noteworthy that our design scheme can realize near-unity total coupling efficiencies even when applied to different material combinations, such as GaAs-based SPSs on $Si_3N_4$ waveguides[34].

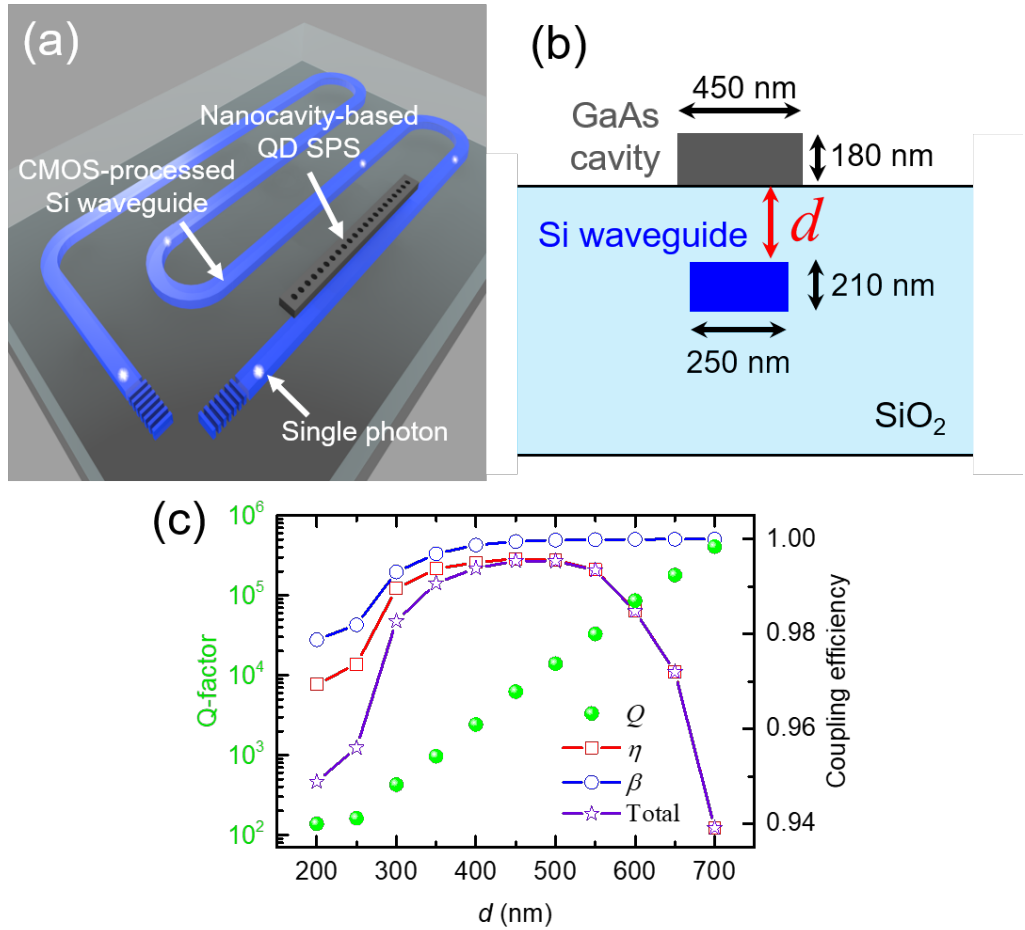

FIG. 1. (a) Schematic of a nanocavity-based QD SPS transfer-printed on a CMOS-processed silicon waveguide with glass cladding. (b) Schematic of the device cross-section. (c) Simulated coupling efficiencies and Q-factors plotted as a function of $d$. Red squares are $\eta$s, blue circles $\beta$s, purple stars total single-photon coupling efficiencies ($\eta\beta$s) and green circles $Q$s. All the simulations discussed in this study are based on the 3D FDTD method.

To fabricate the designed structure, we first prepared InAs/GaAs QD SPSs and silicon waveguides separately. We fabricated nanocavity-based QD SPSs into a 180 nm-thick GaAs slab containing one layer of self-assembled InAs QDs grown by molecular beam epitaxy. We patterned the 1D PhC nanobeam cavities through electron beam lithography and dry and wet etching processes. Figure 2(a) displays a scanning electron microscope (SEM) image of a fabricated nanobeam cavity. Concurrently, we obtained silicon wire waveguides through a CMOS-process foundry. A photograph of a silicon chip is provided in Fig. 2(b). The silicon waveguide is terminated by two output ports to extract waveguide-coupled QD emission into free space directly. The silicon waveguide is buried in a silicon dioxide layer formed by chemical vapor deposition with a thickness of 2 µm. We tuned the thickness of the glass layer above the silicon waveguide (= $d$) to be 350 nm by using a dry etching process. This thickness was chosen to realize high coupling of QD radiation into the waveguide even under the presence of fabrication imperfection.

Next, we used transfer printing to place a fabricated QD SPS onto a CMOS-processed silicon waveguide. Figures 2(c) and (d) show the fabrication procedures for transfer printing. A prepared QD SPS is picked up by placing a polydimethylsiloxane (PDMS) rubber stamp on it and then quickly pealing the rubber stamp off [Fig. 2(c)]. We then place the picked-up QD SPS accurately on top of the silicon waveguide. After putting the SPS, we slowly peel off the rubber stamp, leaving only the QD SPS on the waveguide [Fig. 2(d)]. The QD SPS is firmly bonded on the silicon chip through van der Waals force. These assembling processes are performed with a homemade transfer-printing apparatus composed of precision motion-controlled stages operating under an optical microscope[34]. In the current setup, success rates of both processes of picking-up and placement are nearly 100%. Figure 2(e) shows a microscope image of a completed device. Precise alignment between the top nanobeam cavity and underlying waveguide can be seen. The position deviation between the nanobeam and waveguide is deduced to be < 100 nm, which is routinely possible with our transfer-printing system[34]. We emphasize that the hybrid integration process discussed here is fully compatible with silicon CMOS back-end processes, which is highly advantageous when exploiting the potential of silicon photonics. The transfer-printing approach is also beneficial to integrate multiple QD SPSs: we can sort out proper QD SPSs prior to their assembly onto a target silicon PIC.

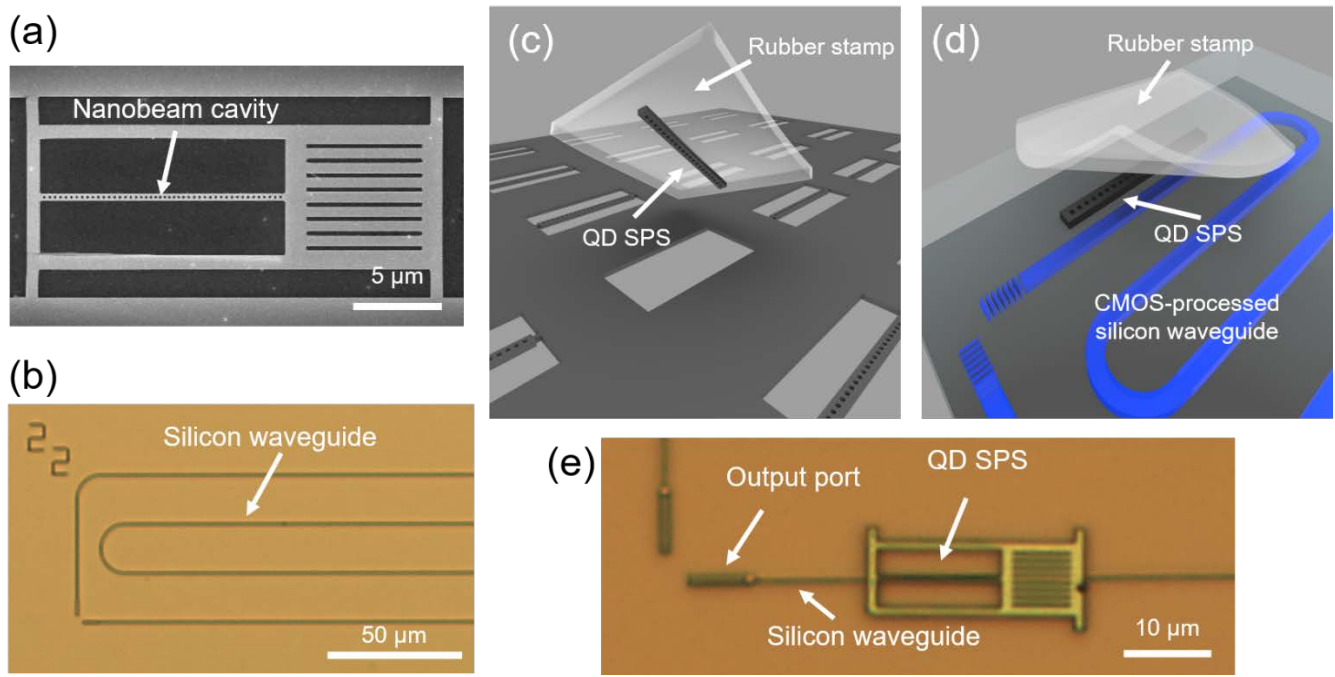

FIG. 2. (a) SEM image of a fabricated nanobeam cavity. (b) Optical microscope image of a silicon wire waveguide obtained through a CMOS-process foundry. (c)-(d) Schematics illustrating the procedure of transfer printing. (e) Microscope image of a completed device.

To characterize fabricated devices, we conducted low-temperature micro-photoluminescence (µPL) measurements. The devices were cooled inside a helium flow cryostat with a built-in heater for controlling the device temperature. We used a x50 objective lens (numerical aperture = 0.65) to image the samples, focus a pump laser beam on the cavity, and collect PL signals. The collected PL was analyzed with a grating spectrometer equipped with an InGaAs camera.

Figure 3(a) displays a visible microscope image of the investigated sample together with a PL image captured at 5 K under the irradiation of a pump laser beam onto the cavity center. Here, we used a continuous wave Ti:sapphire laser oscillating at 820 nm with an average pump power of 13 µW, and spectrally filtered the scattering of the pump laser beam. We observed optical out-coupling from the output ports (the dotted red circles in Fig. 3(a)). These bright signals were not visible when the pump position was shifted from the cavity center. These results confirm that the signals at the output ports primarily stem from the cavity mode radiation coupled to the waveguide.

We then guided PL signal leaked from the upper output port into the spectrometer and measured a PL spectrum, as shown by the red curve in Fig. 3(b). The measured spectrum exhibits a strong peak of cavity mode emission at 1,155 nm, together with that from a QD (labeled as QD-A). For comparison, we measured free space radiation from the sample by collecting PL above the cavity, as plotted by the blue curve in Fig. 3(b). The suppressed cavity signal implies its efficient coupling into the waveguide and marginal radiation into free space.

We estimate an experimental cavity-waveguide coupling efficiency ($\eta_{\text{exp}}$) based on measured cavity Q-factors. From the spectrum in Fig. 3(c), the experimental Q-factor of the cavity being considered is deduced to be $Q_{\text{exp}} = 1,400$. In addition, we evaluate Q-factors of the cavities placed on plane glass and completely decoupled from the silicon waveguide. The average Q-factor ($Q_{\text{ave}}$) estimated from 10 bare nanobeam cavities is estimated to be 14,000. Using these values, we deduce an $\eta_{\text{exp}}$ of 90%[34], elucidating the efficient coupling of the cavity mode into the silicon waveguide.

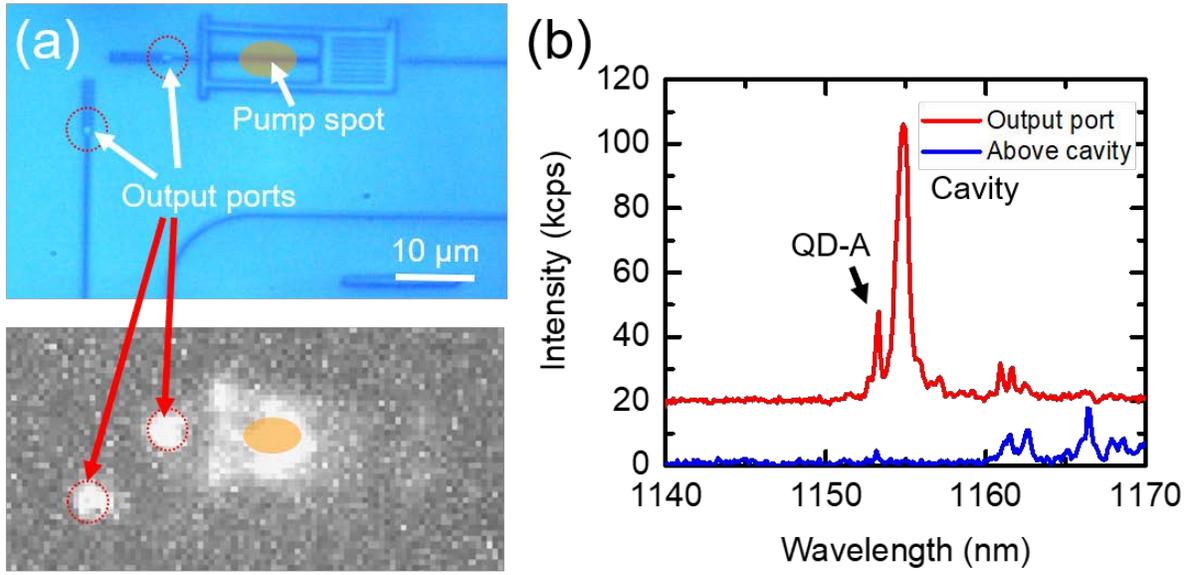

FIG. 3. (a) (Top panel) Visible microscope image of the investigated device. (Bottom panel) PL image of the device while irradiating a pump laser beam onto the cavity center. (b) Measured PL spectra at the upper output port (red curve) and just above the center of the cavity (blue).

Next, we characterized coupling between QD-A and the cavity mode by modifying the energy detuning between them through temperature tuning. Figure 4(a) shows a color plot of PL spectra that were observed through the output port with varying temperatures from 33 K to 52 K. An enhancement of the emission from QD-A was observed when being tuned to the cavity resonance, which suggests that the experimental QD-cavity coupling efficiency ($\beta_{exp}$) is improved as a result of the Purcell effect within the nanocavity. To further confirm the Purcell enhancement in QD-A, we performed time-resolved PL spectroscopy by using a time-correlated single-photon counting technique with a superconducting nanowire single-photon detector (overall system time resolution of ~ 50 ps). For this particular measurement, we switched the pump source to a Ti:sapphire mode-locked laser oscillating at 830 nm (pulse width = ~ 1 ps, repetition rate = 80.3 MHz, average pump power = 180 nW). Figure 4(b) shows a time-resolved PL spectrum measured for the QD-A peak at 40 K, where QD-A is slightly detuned from the cavity resonance by 1.5 nm (shown in the red curve). Here, the spectrometer was used as a band-pass filter for selectively measuring the QD-A peak. The measured curve is fitted with a double exponential decay curve convolved with a function that reflects the system time response. From the red curve in Fig. 4(b), the spontaneous emission rate of QD-A was measured to be 1.8 GHz. The emission rate is found to fasten as the QD-A peak is tuned to the cavity resonance, confirming that the radiative process of QD-A is enhanced by the Purcell effect. For comparison, we measured time-resolved PL spectra for ensemble QD emission observed in an unprocessed area of the same wafer. A typical decay curve is overlaid by a black curve in Fig. 4(b). The bare QDs radiatively decay with an average rate of 0.9 GHz. Based on these experimental results and the fact that a 1D photonic bandgap in our PhC nanobeam typically reduces the emission rate of an embedded QD by half[36], $\beta_{exp}$ can be estimated to be ~ 75%[34,37]. Given $\eta_{exp} = 90\%$, we estimate the total single-photon coupling efficiency, $\eta_{exp}\beta_{exp}$, to be ~ 70%.

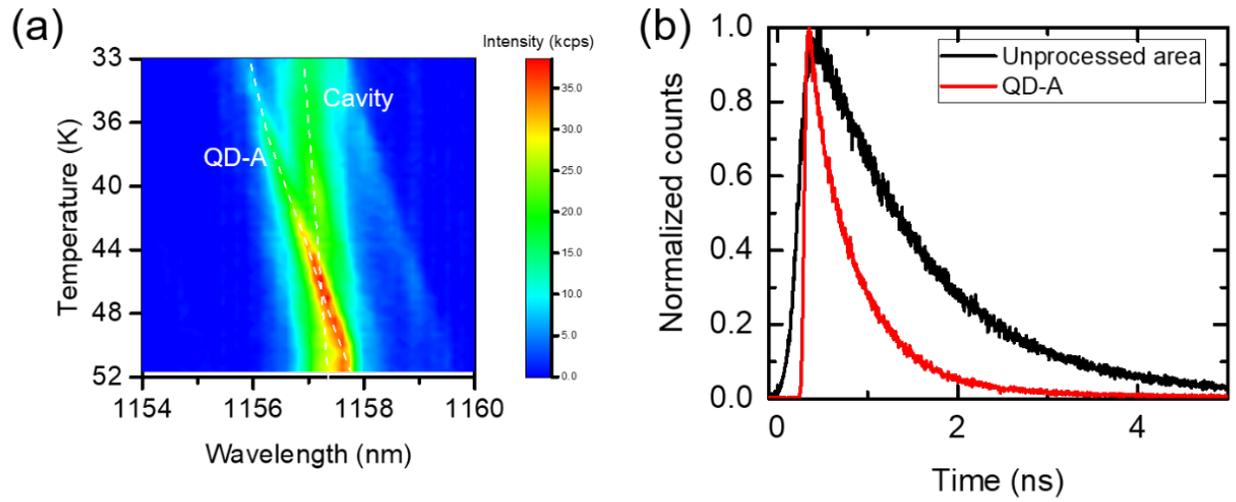

FIG. 4. (a) Color plot of the spectra measured by temperature tuning and through the output port. (b) Time-resolved PL spectra measured for the QD-A emission at 40 K when QD-A is slightly detuned from the cavity resonance by 1.5 nm (red curve). For comparison, we also measured a time-resolved PL spectrum for a typical ensemble QD emission inside an unprocessed area of the same wafer (as shown by the black curve).

Finally, we performed an intensity correlation measurement to confirm the single-photon nature of the emission from QD-A. For this measurement, we added a fiber beam splitter and a superconducting nanowire single-photon detector to the photon counting setup. Figure 5(a) shows a PL spectrum measured at one of the output ports when the cavity center is pumped using a continuous wave laser (wavelength = 860 nm, power = 130 μW). The emission peak of QD-A at 1,153 nm is filtered and sent to the interferometer. Figure 5(b) shows the normalized second-order correlation function $g^{(2)}(t)$ measured for the QD-A peak. The obtained data were fitted with a function, $g^{(2)}(t) = 1 - \left(1 - g^{(2)}(0)\right)\exp(-t/\tau)$, after convolving it with the time response of our detection system, as depicted by the red curve in Fig. 5(b). An anti-bunching with $g^{(2)}(0) = 0.30$ at the zero delay time demonstrates single-photon generation from QD-A. The non-zero $g^{(2)}(0)$ value is likely due to the intrusion of background cavity emission supplied by other off-resonant QDs inside the cavity.

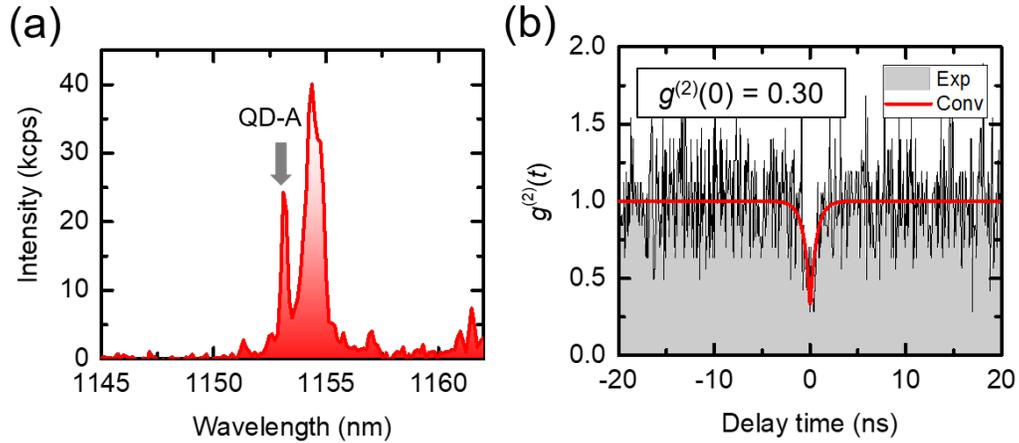

FIG. 5. (a) PL spectrum measured at one of the output ports when the cavity center is pumped using a continuous wave laser at 860 nm. (b) Second-order correlation function $g^{(2)}(t)$ measured for QD-A.

In summary, we demonstrated the hybrid integration of a QD SPS into a CMOS-processed silicon waveguide by transfer printing. We numerically showed that the investigated SPS structure can support near-unity coupling of single photons emitted from the QD into the silicon waveguide. We verified single-photon generation from an integrated QD on silicon and its efficient waveguiding in the CMOS-processed photonic chip. The hybrid integration based on transfer printing can flexibly implement even further nanophotonic components on chip such as nanolasers[38] and cavity quantum electrodynamics systems[39], and therefore enables the synergies of these novel elements and silicon photonics at will.

The authors thank S. Ishida and K. Kuruma for fruitful discussions. This work was supported by JSPS KAKENHI Grant-in-Aid for Specially Promoted Research (15H05700), KAKENHI 16K06294, JST PRESTO (JPMJPR1863) and a project of the New Energy and Industrial Technology Development Organization (NEDO).